# The Feasibility of Asynchronous Rotation via Thermal Tides for Diverse Atmospheric Compositions

Andrea M. Salazar[1]  and Robin Wordsworth[1,2]

[1] Department of Earth and Planetary Sciences, Harvard University, Cambridge, MA 02138, USA; andreasalazar@g.harvard.edu
[2] School of Engineering and Applied Sciences, Harvard University, Cambridge, MA 02138, USA

## Abstract

The equilibrium rotation rate of a planet is determined by the sum of torques acting on its solid body. For planets with atmospheres, the dominant torques are usually the gravitational tide, which acts to slow the planet's rotation rate, and the atmospheric thermal tide, which acts to spin up the planet. Previous work demonstrated that rocky planets with thick atmospheres may produce strong enough thermal tides to avoid tidal locking, but a study of how the strength of the thermal tide depends on atmospheric properties has not been done. In this work, we use a combination of simulations from a global climate model and analytic theory to explore how the thermal tide depends on the shortwave and longwave optical depth of the atmosphere, the surface pressure, and the absorbed stellar radiation. We find that for planets in the habitable zones of M stars only high-pressure but low-opacity atmospheres permit asynchronous rotation owing to the weakening of the thermal tide at high longwave and shortwave optical depths. We conclude that asynchronous rotation may be very unlikely around low-mass stars, which may limit the potential habitability of planets around M stars.

*Unified Astronomy Thesaurus concepts:* Habitable planets (695); Exoplanet atmospheres (487); Atmospheric tides (118)

## 1. Introduction

M stars are the most plentiful star type in the Galaxy, accounting for nearly 75% of stars (N. Reid & S. Hawley 2005). As these stars are much smaller and dimmer than the Sun, the habitable zone (HZ) around an M star is much closer (J. F. Kasting 1997). Strong gravitational tidal forces distort the planet's solid body, leading to a gravitational tidal torque that slows the planet's spin rate. On relatively short timescales, a planet in the HZ of an M star may be expected to become tidally locked (a resonant orbital state where the orbital and rotation rates of the planet are equal; J. F. Kasting et al. 1993; R. Barnes 2017). Such a configuration produces intense gradients in incident stellar radiation, as one side of the planet (the dayside) will be in constant day while the other side (the nightside) will experience constant night.

Such non-Earth-like conditions have prompted decades of study on the atmospheric dynamics and climates of tidally locked planets, using models of complexity spanning global circulation models (GCMs; M. M. Joshi et al. 1997; T. M. Merlis & T. Schneider 2010; J. Yang et al. 2013, 2019; Y. Hu & J. Yang 2014; M. Hammond & N. T. Lewis 2021) to simple models (R. T. Pierrehumbert 2010a; J. Yang & D. S. Abbot 2014; D. D. Koll & D. S. Abbot 2015; R. Wordsworth 2015; W. Kang & R. Wordsworth 2019; D. D. B. Koll 2022). The stark surface temperature gradients on these planets may induce climate states hostile to life as we know it, such as an atmospheric collapse on the nightside (M. M. Joshi et al. 1997; E. S. Kite et al. 2011; R. Wordsworth 2015) or a runaway greenhouse (S. Nakajima et al. 1992). Much work has studied how strong atmospheric heat transport mediated by the large-scale, thermally direct overturning circulation on such slow-rotating planets may mitigate the strong temperature gradients between the dayside and nightside, thus keeping the planet habitable (M. M. Joshi et al. 1997; T. M. Merlis & T. Schneider 2010; R. T. Pierrehumbert & M. Hammond 2019; W. Kang & R. Wordsworth 2019; H. Wu et al. 2023). Additionally, strong ocean heat transport may contribute to efficient heat redistribution (J. Yang et al. 2019). Other work has suggested stabilizing climate feedbacks such as a convective substellar cloud feedback that may enhance the albedo of the dayside, thus maintaining habitable temperatures there (J. Yang et al. 2013).

The potential habitability of tidally locked planets continues to be a question of great interest, especially with the launch of the James Webb Space Telescope (JWST) and the new opportunities it presents to study small, rocky worlds (R. Wordsworth & L. Kreidberg 2022). Recent nondetections of atmospheres on the assumed tidally locked worlds in the TRAPPIST system (S. Zieba et al. 2023; T. P. Greene et al. 2023) may point to a history of atmospheric escape or collapse on the nightside, the latter of which may be connected to the planet's tidally locked state. Understanding the coevolution of atmospheres and orbit states is therefore critical in this new age of atmospheric characterization of exoplanets.

Importantly, the assumption that close-orbiting rocky planets must be tidally locked has been questioned (A. C. Correia & J. Laskar 2003; J. Leconte et al. 2015; P. Auclair-Desrotour et al. 2018). Asynchronous rotation is observed on some planets where synchronous rotation might be expected, most notably Venus, whose slow rotation has long been theorized to be an equilibrium of the gravitational and atmospheric thermal tide (P. Goldreich 1966; T. Gold & S. Soter 1969; A. C. M. Correia & J. Laskar 2001). The time evolution of a planet's rotation rate ($\Omega$) is determined by the sum of torques

$(T_i)$ on the planet's solid body of principle moment of inertia $I$,

$$I\frac{d\Omega}{dt} = \sum_i T_i = T_g + T_a, \qquad (1)$$

where the aforementioned gravitational tide ($T_g$) may not be the only relevant tidal torque enacted on a planet. The atmospheric thermal tide, induced by differential heating and subsequent mass redistribution in a planet's atmosphere across a diurnal cycle, generates a torque ($T_a$) that opposes the gravitational tide (T. Gold & S. Soter [1969](#); D. Cunha et al. [2015](#); P. Auclair-Desrotour et al. [2018](#)). The strength of the thermal tide relative to the gravitational tide therefore determines whether a planet can maintain an asynchronous rotation.

In this study, we assume that the planet is in a perfectly circular orbit with zero obliquity and that the only torques acting on the planet are the gravitational tide and the thermal tide. The theory of the tides is well developed (e.g., W. M. Kaula [1964](#); P. Goldreich [1966](#); C. D. Murray & S. F. Dermott [2000](#); A. C. Correia & J. Laskar [2003](#); M. Efroimsky & J. G. Williams [2009](#); A. C. Correia & J. Laskar [2010](#)), but we summarize the key ideas and equations below for clarity.

### 1.1. The Gravitational Tide

Small differences in the gravitational force from the host star on the solid body of a planet cause a deformation in the planet's shape. This in turn alters the gravitational potential created by the planet. Energy dissipation in the interior of the planet leads to an angular lag between the orientation of the tidal bulge and the planet−star axis (Figure 1, $\delta_g$). The star and planet therefore torque each other, which leads to changes in the planet's rotation rate. The torque on the planet due to this solid-body deformation can be written as

$$T_g = -\frac{3}{2}\frac{GM_*^2 R_p^5}{a^6}b_g(2\Omega - 2n), \qquad (2)$$

where $G$ is the gravitational constant, $M_*$ is the mass of the star, $R_p$ is the radius of the planet, $a$ is the semimajor axis of the planet's orbit around the star, and $n$ is the mean orbital rate of the planet ($2\pi/T_{\rm orb}$, with $T_{\rm orb}$ the orbital period. The high power dependence of the gravitational tide on $a$ means that planets in the HZ of small stars such as M stars experience very strong gravitational tidal effects compared to HZ planets around Sun-like stars. Similarly, the gravitational tide is very strong for large planets owing to the $R_p^5$ dependence. The term $b_g(2\Omega - 2n)$ is the tidal response of the solid body, describing both the amplitude of the tide and the lag angle between the orientation of the tidal bulge and the planet−star axis. In general, the term $b_g(2\Omega - 2n)$ depends on both the rheology of the planet and the forcing frequency ($\sigma = \Omega - n$). In this work we will use the constant-$Q$ model (W. M. Kaula [1964](#)) for simplicity, for which

$$b_g(2\Omega - 2n) = \frac{k_2}{Q}\,{\rm sgn}(2\Omega - 2n), \qquad (3)$$

where $k_2$ is the second-order tidal Love number and $Q$ is the quality factor, describing energy dissipation in the interior. This is equivalent to assuming that the lag angle, $\delta_g$, is independent of planetary rotation rate. The validity of the constant-$Q$ model has been challenged (M. Efroimsky & V. V. Makarov [2013](#)), but our aim is to compare the relative magnitudes of the gravitational and thermal tides, not accurately predict the spin evolution of the planet, so we utilize the constant-$Q$ model to allow us to derive analytic formulae. Given this assumption, the torque due to the gravitational tide is given by

$$T_g = -\frac{3}{2}K_g\,{\rm sgn}(2\Omega - 2n), \qquad (4)$$

where $K_g = GM_*^2 R_p^5 k_2/Qa^6$ with units of N·m. For a planet rotating supersynchronously ($\Omega > n$), the gravitational tide will spin down the planet. For a planet rotating subsynchronously ($\Omega < n$), the gravitational tide will spin up the planet. If no other torques were acting on the planet, the equilibrium state would be a synchronous rotation ($\Omega = n$), and the planet would be tidally locked.

### 1.2. The Atmospheric Thermal Tide

The atmospheric thermal tide is caused by large-scale mass redistribution of a planet's atmosphere due to differential stellar heating over the diurnal cycle (R. S. Lindzen [1967](#)). As for the gravitational tide, the mass redistribution raises a tidal potential that is torqued by the star. The torque on the atmosphere is then communicated to the solid body via boundary layer friction (A. C. Correia & J. Laskar [2003](#)).

We derive the tidal potential, $V_a$, induced by mass redistribution (see J. Leconte et al. [2015](#) for a comprehensive derivation). We assume that the atmosphere is in hydrostatic balance such that the vertical pressure (and therefore mass) profile can be calculated from the surface pressure. Then,

$$V_a(\mathbf{r}) = -\frac{GR_p}{g}\sum_{l=0}^{\infty}\left(\frac{R_p}{r}\right)^{l+1}\frac{4\pi}{2l+1}\sum_{m=-l}^{l}\mathbf{Y}_l^m(\theta, \phi)$$
$$\times \int p_s(\theta', \phi')Y_l^{m*}(\theta', \phi')\sin\theta' d\theta' d\phi', \qquad (5)$$

where we have expanded the Newtonian potential into the spherical harmonics, $Y_l^m(\theta, \phi)$, where $\theta$ is the polar angle and $\phi$ is the azimuthal angle. The integral in Equation ([5](#)) performs a spherical harmonic decomposition of the surface pressure field, $p_s(\theta, \phi)$. The torque of the atmosphere on the star (and, equivalently, the equal and opposite torque of the star on the atmosphere) for a planet with zero obliquity in a perfectly circular orbit is $T_a = M_* r\left(\frac{1}{r}\frac{\partial}{\partial\phi}V_a\right)$.

The $l = 0$ and $l = 1$ terms do not contribute to the torque, as they correspond to either a perfect sphere or a center-of-mass shift, neither of which causes torque (A. C. Correia & J. Laskar [2003](#)). In addition, the torque scales like $(R_p/r)^{l+1}$, and since $R_p/r$ is very small when $r$ is taken to be the semimajor axis of the planet's orbit, $a$, terms higher than $l = 2$ contribute very little to the torque. Therefore, we retain only the $l = 2$ term, the quadrupole moment of the surface pressure. Because we have assumed a planet with zero obliquity such that the pressure field is assumed to be symmetric about the equator, we define our coordinate system at the equator ($\theta = \pi/2$), with longitude measured from the substellar point. After some algebra, we can write the torque due to the

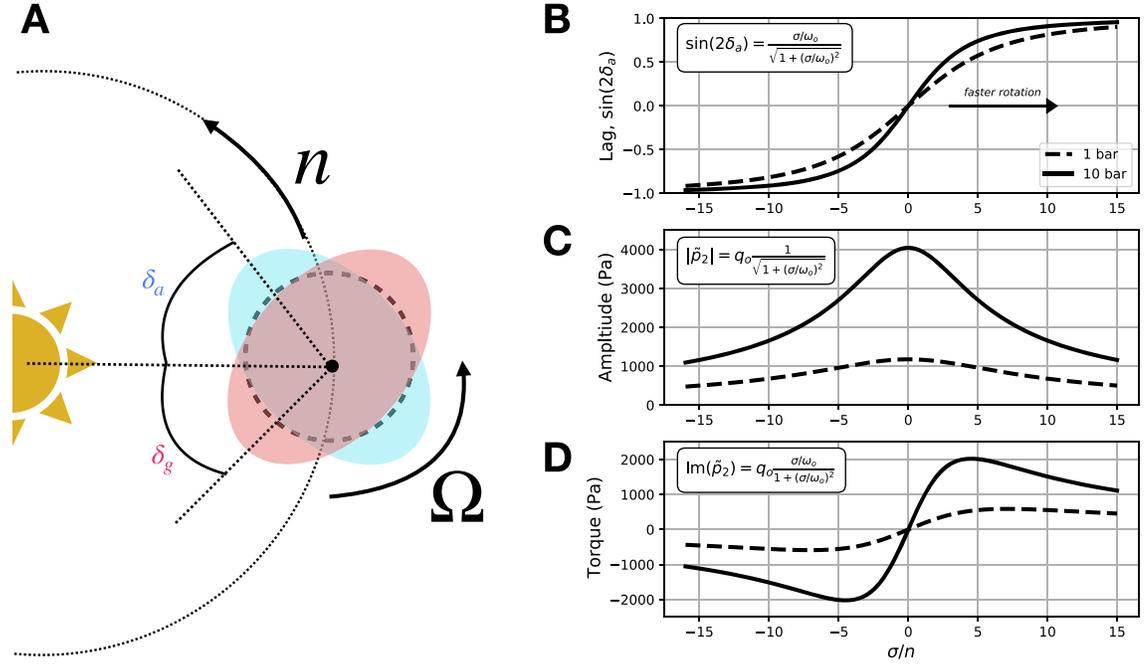

**Figure 1.** (a) Schematic demonstrating the amplitude and lag angles of the gravitational tide and the atmospheric thermal tide. The lag angle is measured from the planet−star axis and is given by $\delta_g$ for the gravitational tide and $\delta_a$ for the atmospheric thermal tide. As long as the pressure bulge (the blue ellipsoid) leads the planet−star axis, the thermal tide will produce a torque that opposes the gravitational tide. (b–d) Summary of results from J. Leconte et al. (2015), demonstrating the forcing frequency, $\sigma = \Omega - n$, dependence of the thermal tide, decomposed into (b) the lag angle, (c) the amplitude, and (d) the torque due to the thermal tide. Inset boxes display the functional form of each component, derived from their simple model of a periodically heated thermal slab. A forcing frequency of zero corresponds to a tidally locked case where the rotation rate, $\Omega$, is equal to the mean orbital rate, $n$.

atmospheric tide as

$$T_a = -\frac{4\pi}{5}\frac{GM_*R_p}{g}\left(\frac{R_p}{a}\right)^3 \sum_{m=-2}^{m=2} \tilde{p}_2^m \frac{\partial}{\partial\phi} Y_2^m\left(\frac{\pi}{2}, \phi\right),$$

where $\tilde{p}_2^m \equiv \int p_s\left(\frac{\pi}{2}, \phi'\right) Y_2^{m*}\left(\frac{\pi}{2}, \phi'\right) d\phi$. The term $Y_2^m\left(\frac{\pi}{2}, \phi\right)$ is zero for $l = -1, 0, 1$ and $Y_2^2 = Y_2^{2*}$. If we define the substellar point as the origin of longitudes (meaning that the surface pressure field must also be in this reference frame), then with a bit more algebra the torque due to the thermal tide is

$$T_a = \frac{3}{2}K_a\,\mathrm{Im}(\tilde{p}_2), \tag{6}$$

where $K_a = \sqrt{\frac{12}{10\pi}}\left(M_*R_p^3/\bar{\rho}a^3\right)$ with units of m$^3$ and $\mathrm{Im}\,(\tilde{p}_2) \equiv \mathrm{Im}\left(\int p_s\left(\frac{\pi}{2}, \phi'\right)Y_2^{2*}\left(\frac{\pi}{2}, \phi'\right)d\phi\right)$ is the imaginary part of the quadrupole moment of the pressure field (light-blue ellipsoid in Figure 1(a)) with units of Pa. Therefore, Equation (6) has units Pa·m$^3$, or N·m. To calculate the torque due to the thermal tide in this formulation, all that is required is the surface pressure field at the equator, with the substellar point defined as the origin of longitudes. We have also defined the average bulk density of the planet, $\bar{\rho}$. Note that the thermal tide has an $a^{-3}$ dependence, smaller than the $a^{-6}$ dependence of the gravitational tide. This means that HZ planets will require stronger thermal tides around smaller stars to avoid synchronization.

As in the gravitational tide, the torque due to the thermal tide is determined by both the amplitude of the surface pressure quadrupole, $|\tilde{p}_2|$, and the lag angle ($\delta_a$). The atmosphere has thermal inertia and does not respond to stellar forcing instantaneously. Therefore, the hottest part of the atmosphere is not necessarily at the substellar point, so the pressure bulge is not perfectly aligned with the planet−star axis. This is a critically important effect because this misalignment is what generates the torque. In the tidally locked case, the amplitude of the pressure quadrupole is quite large, but because the lag angle is zero (as the substellar point is fixed in this case), the torque is zero.

The relevance of the atmospheric thermal tide to rocky planet spin evolution was first explored for Venus, the retrograde, subsynchronous rotation of which appears to be an equilibrium of the gravitational and thermal tide (T. Gold & S. Soter 1969; A. R. Dobrovolskis & A. P. Ingersoll 1980; A. C. M. Correia et al. 2003). More recently, the atmospheric thermal tide has been used to understand changes in Earth's spin rate throughout its history (H. Wu et al. 2023a; M. Farhat et al. 2024). From these examples in the solar system, it is clear that the atmospheric thermal tide has a meaningful control on the rotational evolution of rocky HZ planets and therefore should be studied in depth for applications to rocky exoplanets.

Previous work by J. Leconte et al. (2015) used the Laboratoie de Meteorologie Dynamique (LMD) Generic Model

to study the forcing frequency (or, similarly, the rotation rate) dependence of the thermal tide, which they found to be well approximated by a simple model of a periodically heated thermal slab (Figure 1(d)). The torque due to the thermal tide varies nonmonotonically with rotation rate, because faster rotations both increase the lag angle (Figure 1(b)) and decrease the amplitude of the thermal tide (Figure 1(c)). The thermal tide thus behaves like a forced oscillator with some internal characteristic frequency (in this case, the thermal equilibrium frequency, $\omega_o$, which essentially represents the radiative response time of the atmosphere). This results in a maximum in the strength of the thermal tide at forcing frequencies, $\sigma$, similar to the thermal equilibrium frequency of the atmosphere. All else being equal, the lag angle increases with rotation rate because the longitude of the substellar point moves faster than the atmosphere can thermally adjust, leading to a larger lag between the orientation of the complex quadrupole and the planet−star axis. This would tend to strengthen the magnitude of the torque. However, increasing the rotation rate also decreases the amplitude of the thermal tide because as the length of a day decreases temperature anomalies become smaller. The competition of these two effects leads to the nonmonotonic dependence of torque on rotation rate.

J. Leconte et al. (2015) also found that the thermal tide strengthens with surface pressure (Figure 1(d), solid line vs. dashed line) and suggested that planets with thick atmospheres (around 10 bars) could have thermal tides strong enough to oppose the gravitational tide around M star planets in the HZ, therefore maintaining asynchronous rotation. In each GCM study, they assumed either cloud-free Earth-like atmospheric compositions or pure $CO_2$. Though their simple thermal slab model was a powerful predictor of the frequency dependence of the thermal tide, it required the input of key parameters (the amplitude of the pressure quadrupole at zero frequency, $q_o$, and the thermal equilibrium frequency, $\omega_o$) from the GCM, necessitating a GCM run for every atmospheric composition considered. This precluded a wide parameter space survey of how atmospheric properties affected the strength of the thermal tide.

Due to the stochastic nature of planetary formation and atmospheric loss, we can expect terrestrial planets to have a wide range of atmospheric compositions, and therefore studies of the thermal tide should explore atmospheres different from Earth's. Additionally, atmospheres evolve and change composition throughout a planet's history, which may result in previously unexplored climate−rotation feedbacks that may affect the likelihood of asynchronous rotation. In this work, we seek to develop a more holistic understanding of how atmospheric composition affects the strength of the thermal tide to determine how likely asynchronous rotation is for a diverse set of atmospheres. We do this using a hierarchical modeling approach, using 3D GCMs and analytic models to develop intuition and theory that describe how the strength of the thermal tide depends on atmospheric properties. This allows us to map out the parameter regimes under which we might expect asynchronous rotation of rocky exoplanets.

In Section 2, we perform targeted GCM runs about a base, "Earth-like" atmospheric state to gain intuition on the effect of atmospheric properties on the strength of the thermal tide. In Section 3, we develop a simple model that predicts surface pressure anomalies and thermal tidal torques. We further simplify the model by performing a linearization in Section 3.1, which allows us to write an analytic expression for the torque and the equilibrium rotation rate as a function of various atmospheric and planetary properties in agreement with the GCM results. In Section 4, we use the analytic model to map out the conditions under which a planet may escape tidal locking, and we show that there is only a narrow range of parameters for which atmospheres are capable of producing strong enough thermal tides around small stars. We conclude in Section 5 with a discussion of the broader implications of our work.

## 2. GCM Results

For the most consistent follow-up study to J. Leconte et al. (2015), we use the LMD Generic Model (F. Forget et al. 2013; R. Wordsworth et al. 2013). The LMD model solves the primitive equations on a sphere with a spatial longitude −latitude−altitude resolution of $64 \times 48 \times 18$ and scaled $\sigma$-pressure coordinates in the vertical (R. Wordsworth et al. 2013). We run the model with both gray-gas radiative transfer and real-gas radiative transfer, the latter of which uses a correlated-$k$ approach (R. Wordsworth et al. 2010). To allow a more direct comparison with the results of J. Leconte et al. (2015), we choose the same orbital configuration, with an orbital period of 224 days. In all of our simulations, the planet is one Earth radius, has a uniform albedo of 0.2, and has no liquid water or clouds. We set a planetary rotation rate equal to two times the orbital rate such that the planet is in a 2:1 spin–orbit resonance. We chose this spin state so that the planet would be firmly in the slow-rotation regime while also allowing us to explore how changes to the lag angle with varying atmospheric composition contribute to the torque. Had we chosen a tidally locked state, the lag angle (and therefore the torque) would always be zero. The slow rotation rate chosen for the GCM study also allows exploration of how atmospheric composition affects the torque in a regime where the lag angle has not saturated at 90°, as often occurs at faster rotation rates.

We investigate the sensitivity of the thermal tide to atmospheric properties by performing a series of targeted sensitivity tests in the gray-gas radiative transfer configuration, in which we independently vary four parameters about a base state: longwave optical depth of the atmosphere ($\tau_{LW}$), shortwave optical depth of the atmosphere ($\tau_{SW}$), absorbed shortwave radiation (ASR), and atmospheric surface pressure ($p_s$). Table 1 details the parameter ranges explored in this study, as well as the base state of each parameter. We choose the base state to be approximately "Earth-like" with an incident shortwave radiation (ISR) approximately in the center of the HZ (R. K. Kopparapu et al. 2013). We then select parameter ranges to encapsulate a parameter space spanning a wide diversity of terrestrial exoplanet atmospheres. We set upper limits on the surface pressure and ASR based on Venus's approximate surface pressure and incident stellar radiation. The upper limits of $\tau_{LW}$ and $\tau_{SW}$ were both set by model nonconvergence at higher values. This parameter space is by no means exhaustive but allows us to investigate the sensitivity of the strength of the thermal tide to atmospheric composition.

In Figure 2, we show the resulting sine of the lag angle, amplitude, and torque of the atmospheric thermal tide from the GCM with varying $\tau_{LW}$, $\tau_{SW}$, ASR, and $p_s$. The optical depths, $\tau_{LW}$ and $\tau_{SW}$, are declared directly by changing the absorption coefficient of the atmosphere in either the longwave or shortwave. ASR = ISR(1-$\alpha$) is set directly and independently

**Table 1**
Parameters Varied in GCM Sensitivity Test, Including Reference Values and Value Ranges

| Parameter | Reference Value | Value Range |
|---|---|---|
| Atmospheric surface pressure, $p_s$ (bars) | 1[*] | 0.1–100 |
| Longwave optical depth of the atmosphere, $\tau_{LW}$ | 1 | $10^{-4}$–$10^2$ |
| Shortwave optical depth of the atmosphere, $\tau_{SW}$ | $10^{-4}$ | $10^{-4}$–10 |
| Incident shortwave radiation ISR (W m$^{-2}$) | 910 | 100–2600 |
| Orbital mean motion, $n$ | $2\pi/224$ days[*] | ⋯ |
| Rotation frequency, $\Omega$ | $2\pi/112$ days | ⋯ |
| Orbital semimajor axis, $a$ (AU) | 0.72[*] | ⋯ |
| Surface gravity, $g$ (ms$^{-2}$) | 9.8[*] | ⋯ |
| Surface albedo, $\alpha$ | 0.2[*] | ⋯ |
| Soil thermal inertia (tiu) | 270 | ⋯ |
| Diffusivity factor, $D$ | 1.66 | |
| Heat capacity of surface, $C_s$ (J kg$^{-1}$) | $1 \times 10^6$ | |
| Specific heat of air, $c_p$ (J kg$^{-1}$ K$^{-1}$) | 1000 | |
| Reference thermal coupling factor, $\chi$ | 0.17 | |
| Surface drag coefficient, $C_D$ | 0.0034 | |
| Specific gas constant, $R$ (J kg$^{-1}$ K$^{-1}$) | 188 | |
| Reference surface wind speed $U_{so}$ (m s$^{-1}$) | 20.14 | |

**Note.** Parameters are varied one at a time and set to their reference value when a sensitivity test on another parameter is performed. Parameters fixed across all GCM sensitivity tests (middle). Parameters used in the analytic model (bottom) are fixed unless otherwise noted in the text. Stars indicate values that were chosen to match the experimental setup by J. Leconte et al. (2015). We chose the range in ISR to span up to Venus's incident stellar flux, with the reference value about in the center of the HZ range proposed by R. K. Kopparapu et al. (2013).

of orbital distance. When changing pressure, we also adjust the longwave and shortwave absorption coefficients to maintain constant $\tau_{LW}$ and $\tau_{SW}$. The torque (plotted here as the imaginary component of the complex pressure quadrupole) is the product of the amplitude and the sine of the lag angle, but we show all three terms for clarity. The amplitude is simply the norm of the complex pressure quadrupole, while the sine of the lag angle is the argument of the complex pressure quadrupole (J. Leconte et al. 2015). For each simulation, we run the GCM for 10 Earth years. To construct the pressure field, we shift the surface pressure field (output every 2 Earth days) so that the substellar point is located at 0° longitude at every time step. We then average over the last 450 Earth days (about equal to 2 yr for this planet). Our results are insensitive to the length of time averaged over, except in the very high $\tau_{SW}$ regime, discussed below. As the thermal tide is driven by variations in surface temperature, our initial expectation in the sensitivity study was monotonic trends in torque that scale with longitudinal temperature anomalies.

The response of the thermal tide to an increase in shortwave optical depth (Figure 2, green) is most straightforward. Increasing atmospheric absorption of shortwave monotonically weakens the thermal tide by both limiting stellar radiation incident on the surface and stabilizing the atmosphere against large-scale circulation (W. Kang & R. Wordsworth 2019). This leads to nearly an order-of-magnitude decrease in the strength of the thermal tide between the transparent case and the $\tau_{SW} = 2$ case. At $\tau_{SW} = 10$, the torque is essentially zero. The lag angle shows a sharp jump between shortwave optical depth of 2 and 10, but we suspect this to be a result of internal model variability, as the pressure field is quite flat at such high optical depth. Unlike in the other simulations, the lag angle in this regime is very sensitive to the averaging time period. From one time step to another in the $\tau_{SW} = 10$ case, the lag angle is anywhere from −1 to 1. Averaging over the last 500 Earth days yields a lag angle of 0.44 (as seen in Figure 2), whereas averaging over 100 Earth days yields a lag angle of −0.9. In any case, the amplitude is near zero at $\tau_{SW} = 10$, so the lag angle contributes very little to the overall torque.

From the $\tau_{SW}$ sensitivity test, we might expect a similar reduction in the strength of the thermal tide when ASR (=ISR (1-α)) is decreased (either by decreasing the stellar constant or by increasing planetary albedo) in Figure 2 (purple). However, we find that the magnitude of the torque is roughly constant in this range of ASR. Decreasing ASR does decrease the amplitude of the thermal tide by reducing the stellar forcing; however, this leads to a cooler planet with a longer thermal equilibrium timescale ($\omega_o \propto 4\sigma_{SB}T^3$), which increases the lag angle. The result is a complete cancellation of these opposing effects and a nearly constant torque at a rotation rate equal to $2n$. We expect, however, that at rotation rates farther from synchronous rotation the compensation of the amplitude with the lag angle will break down. Lag angle increases with increasing rotation rate and saturates at $\sin(2\delta) = 1$, or, equivalently, $2\delta = 90°$. In this regime, the magnitude of the torque is more sensitive to changes in amplitude, so we expect that we should then recover an increase in torque with increased stellar absorption. This effect will be explored in depth with the simple model in Section 3.1.

An especially interesting result of the GCM sensitivity test is the nonmonotonic behavior of the torque with increasing longwave optical depth. Ignoring all atmospheric dynamics, we might expect that increasing the longwave optical depth of the atmosphere would weaken the thermal tide by decreasing temperature gradients across the dayside and nightside, therefore mitigating pressure gradients. However, we see in Figure 2 (red) that optically thin atmospheres have extremely weak thermal tides, despite large zonal temperature gradients. Increasing the longwave optical depth initially strengthens the thermal tide until around $\tau_{LW} = 1$, where the trend reverses. This nonmonotonic behavior emerges owing to two competing effects with increasing longwave optical depth: the strengthening of large-scale atmospheric circulation and the weakening of zonal temperature gradients.

The thermally direct large-scale overturning circulation on slow-rotating planets has been studied extensively (M. M. Joshi et al. 1997; S. Wang & J. Yang 2022). Longwave cooling on the nightside causes subsidence, which invigorates the circulation and, through mass conservation, requires rising motion on the dayside. GCM studies demonstrated that the addition of even a few parts per billion of $CO_2$ into an otherwise transparent atmosphere causes a strong increase in the strength of the overturning circulation (S. Wang & J. Yang 2022), and the circulation continues to strengthen with increased $CO_2$ content. Similarly, we examined the magnitude of upward vertical velocity on the dayside in the LMD model and found nearly a 10-fold increase across the range of $\tau_{LW}$ examined here. In the optically thin regime, fractional temperature gradients do not decrease much with increased optical depth, and so the behavior of torque is dominated by the strengthening of atmospheric circulation (and therefore increased mass redistribution). Conversely, in the optically thick limit, the behavior of the torque is dominated by the rapid reduction in fractional temperature gradients. The response of the large-scale circulation to parameter changes is therefore of first-order importance to the study of thermal tides.

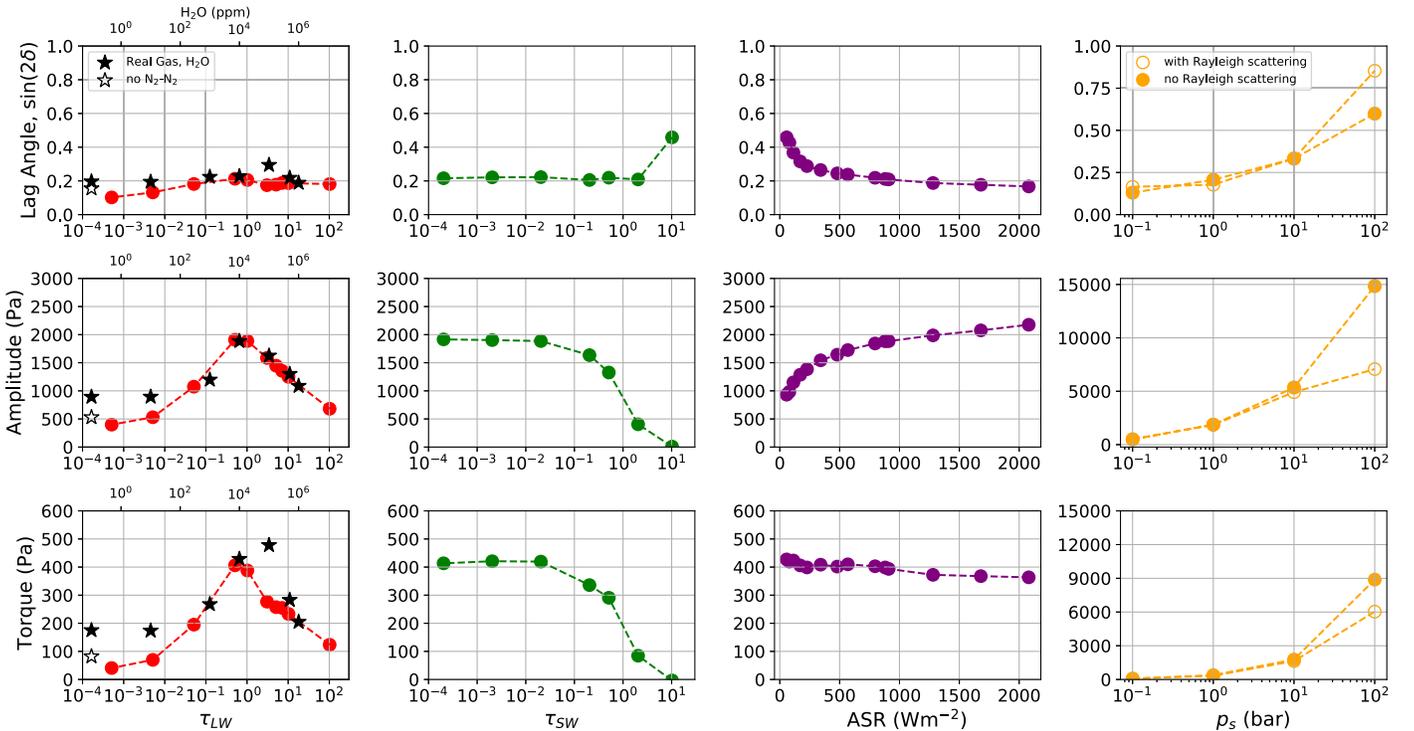

**Figure 2.** Resulting sin of the lag angle (first row), amplitude (second row), and torque (amplitude times lag; third row) of the thermal tide for varying longwave optical depth ($\tau_{LW}$; red), shortwave optical depth ($\tau_{SW}$; green), ASR (purple), and surface pressure ($p_s$; orange). In the first column, black stars in the longwave optical depth test indicate real-gas experiments run with increasing $H_2O$ content in an $N_2$ atmosphere. The open black star demonstrates the effect of disabling $N_2$–$N_2$ collisional absorption. In the last column, open orange circles indicate simulations with Rayleigh scattering. Closed circles indicate simulations where Rayleigh scattering is disabled. In all sensitivity tests, the rotation rate is set to $2n$.

To confirm that this behavior is robust, we performed additional real-gas simulations using the correlated-$k$ method (R. Goody et al. 1989; R. Wordsworth et al. 2010). The black stars in Figure 2 correspond to experiments of an $N_2$-dominated atmosphere with varying prescribed concentrations of water vapor. We do not allow the water to condense onto the surface or form clouds. From the real-gas experiment, we observe the same behavior observed in the gray-gas experiment: an increase in torque until $10^4$ ppm of $H_2O$ followed by a decrease in torque. At low concentrations of $H_2O$, the torque is about twice as high as the optically thin gray-gas case. This is due to the collision-induced absorption of $N_2$–$N_2$. To demonstrate this effect, we disable the $N_2$–$N_2$ absorption in the lowest $H_2O$ simulation (open black star in the first column of Figure 2), which leads to a torque much closer to the torque in the low gray-gas $\tau_{LW}$ case.

A key takeaway by J. Leconte et al. (2015) was that planets with high enough surface pressures could potentially escape tidal locking. The authors suggested an upper limit to this effect owing to the enhancement of shortwave absorption of stellar radiation. In Figure 2 (orange), we vary surface pressure but hold the opacity of the atmosphere constant. As expected, the thermal tide strengthens with increased surface pressure, through both an increase in the amplitude of the pressure bulge and an increase in the lag angle. All things being equal, one might expect a 10-fold increase to lead to a 10-fold increase in the pressure amplitude, but Figure 2 shows approximately a threefold enhancement. This is partly due to a reduction in temperature anomalies with increased surface pressure and a weakening of overturning circulation due to enhanced frictional dissipation at the surface. We also investigate the effect of enhanced Rayleigh scattering by performing surface pressure sensitivity tests where we disable Rayleigh scattering (closed orange circles) and where we allow Rayleigh scattering to increase with surface pressure (open orange circles). Rayleigh scattering acts to increase the albedo of the planet by scattering incoming stellar radiation. A higher surface pressure enhances Rayleigh scattering owing to the increase in the number of molecules in the atmosphere. Therefore, planets with higher surface pressures have lower ASR. In fact, ASR decreases exponentially with surface pressure in this model, ranging from about 720 to 40 W m$^{-2}$ between the 0.1-bar and 100-bar cases. We see that the effect of enhanced Rayleigh scattering appears between 10 and 100 bars and leads to a 30% weaker torque. This is consistent with the ASR study, which showed large increases in lag angle and large decreases in amplitude when ASR dropped below about 500 W m$^{-2}$, which is the ASR in the 10-bar case.

We chose not to explore varying the rotation rate using the GCM, as this was already done by J. Leconte et al. (2015). We note, however, that changes to the rotation rate may affect the response of the torque to changes in certain parameters. We expect this to be especially true in cases where the behavior of the torque is driven by changes in both the amplitude and lag angle. In the ASR sensitivity test, for example, the nearly constant torque was caused by the compensation of a decreasing lag angle and increasing amplitude with increased ASR. At different rotation rates, these two effects will likely not perfectly compensate, which would lead to a different behavior of the torque. We will explore this effect in more depth with the simple model.

These targeted sensitivity tests suggest that the strength of the thermal tide is quite sensitive to atmospheric properties, which may limit the parameter space in which the thermal tide can provide adequate torque to oppose the gravitational tide. To further explore this problem and to gain deeper intuition, we develop a simple model of the atmospheric thermal tide.

## 3. Simple Model

The thermal slab model of J. Leconte et al. (2015) demonstrated that the frequency dependence of the thermal tide is well approximated by a simple model of a periodically heated slab; however, their model required the input of key parameters from GCM runs. In this section, we develop a simple model that predicts the torque due to the thermal tide without necessitating input from GCMs for every atmospheric composition considered. The model's simplicity allows an extensive, computationally efficient exploration of the parameter space.

For terrestrial atmospheres, the thermal tide is primarily driven by heating at the surface, where atmospheric pressure is highest. We therefore write a surface energy balance to calculate the temperature at the equator in time for every longitude on the planet, $\lambda$,

$$C\frac{dT(\lambda)}{dt} = S_o(1 - \alpha(\lambda))e^{-\tau_{SW}}f(\lambda, t) - \sigma_{SB}T(\lambda)^4 + \mathbf{F}_{LW}^-(\tau_{LW}), \tag{7}$$

where $C$ is the heat capacity of the surface plus the boundary layer that is thermally coupled to the surface. $S_o$ is the stellar constant, $\alpha(\lambda)$ is the albedo of the planet at each longitude, and $\tau_{SW}$ is the shortwave optical depth of the atmosphere. For simplicity, we assume that the albedo is constant in longitude. The function $f(\lambda, t)$ describes the spatial and temporal dependence of the instellation

$$f(\lambda, t) = \begin{cases} \cos(\zeta) & -\pi/2 \leqslant \zeta \leqslant \pi/2 \\ 0 & \pi/2 < \zeta < 3\pi/2, \end{cases} \tag{8}$$

where $\zeta = \lambda - \sigma t$ is the stellar zenith angle and $\sigma = \Omega - n$ is the forcing frequency. We denote the Stefan–Boltzmann constant as $\sigma_{SB}$ to avoid confusion with the forcing frequency. To calculate the downwelling longwave flux at the surface, $\mathbf{F}_{LW}^-(\tau_{LW})$, we assume that the atmosphere satisfies the weak temperature gradient (WTG) approximation and is in radiative equilibrium. The WTG approximation has been shown to represent the atmosphere of slow-rotating planets well in GCMs (T. M. Merlis & T. Schneider 2010; R. Wordsworth 2015; D. D. B. Koll & D. S. Abbot 2016) and is employed here for simplicity. The temperature profile is deduced by solving the two-stream equations for an atmosphere with both thermal and stellar absorption following T. D. Robinson & D. C. Catling (2012). The two-stream equations for both the longwave and shortwave flux are (R. T. Pierrehumbert 2010b)

$$\frac{dF_{LW}^+}{d\tau^I} = D[F_{LW}^+ - \sigma_{SB}T(\tau^I)^4] \tag{9}$$

$$\frac{dF_{LW}^-}{d\tau^I} = -D[F_{LW}^- - \sigma_{SB}T(\tau^I)^4] \tag{10}$$

$$\frac{d\mathbf{F}_{SW}^-}{d\tau^V} = F_{SW}^-, \tag{11}$$

where $\tau^I$ is the vertical coordinate corresponding to longwave optical depth and $\tau^V$ is the vertical coordinate corresponding to shortwave optical depth. The upwelling (superscript $+$) and downwelling (superscript $-$) of shortwave and longwave fluxes are $F_{SW}^-$, $F_{LW}^+$, and $F_{LW}^-$, and $D$ is the diffusivity factor, set to 1.66 in this work.

Assuming radiative equilibrium at each level, the heating rate is

$$Q = \frac{d}{d\tau^I}(F_{LW}^+ - F_{LW}^- - F_{SW}^-) = 0. \tag{12}$$

Imposing the boundary condition that at the top of the atmosphere the outgoing longwave radiation (OLR) equals the ASR, such that OLR = ASR = $S_o(1 - \alpha)$, then

$$F_{LW}^+ - F_{LW}^- = F_{SW}^-. \tag{13}$$

Manipulation of Equations (9)–(10) gives

$$\frac{d^2}{(d\tau^I)^2}(F_{LW}^+ - F_{LW}^-) - D^2(F_{LW}^+ - F_{LW}^-) = -2\frac{d}{d\tau^I}(\sigma_{SB}T(\tau^I)^4), \tag{14}$$

which can be solved for $\sigma_{SB}T^4$ to obtain the radiative equilibrium temperature profile. Solving Equation (11),

$$\mathbf{F}_{SW}^-(\tau^I) = \text{ASR}e^{-k\tau^I}, \tag{15}$$

where we have defined $k = \tau_{SW}/\tau_{LW}$. Combining Equations (13)–(15) and integrating over $\tau^I$ gives the radiative equilibrium temperature profile

$$\sigma_{SB}T(\tau^I)^4 = \frac{S_o(1 - \alpha)}{2}[1 - Dk^{-1} + (kD^{-1} - Dk^{-1})e^{-k\tau^I}]. \tag{16}$$

For large values of $k$ corresponding to an atmosphere with strong stellar absorption, a temperature inversion occurs at high altitudes. For small values of $k$, the solution approaches a no-stellar-absorption radiative equilibrium.

Integrating from the top of the atmosphere to the surface, the downwelling longwave at the surface, $\mathbf{F}_{LW}^-(\tau_{LW})$, is given by (T. D. Robinson & D. C. Catling 2012)

$$\mathbf{F}_{LW}^-(\tau_{LW}) = \frac{S_o(1 - \alpha)}{2}[1 + Dk^{-1} - (1 + Dk^{-1})e^{-k\tau_{LW}}]. \tag{17}$$

Finally, we define the heat capacity, $C$, as the sum of the heat capacity of the surface, $C_s$, and the bottom layer of the atmosphere, $C_a$, that is thermally coupled to the surface. Therefore,

$$C = C_s + C_a = C_s + c_p\Delta p/g, \tag{18}$$

where $c_p$ is the specific heat of air, $g$ is the acceleration due to gravity, and $\Delta p$ is the pressure thickness of the portion of the atmosphere that is thermally coupled to the surface (defined below).

One method for solving Equation (7) is to find the time-dependent equatorial temperature oscillation at each longitude and then shift each curve such that $\lambda = 0$ is defined at the substellar point. This phase folding of the solutions constructs a function of temperature with longitude from the substellar point, $\lambda^*$, which is the reference frame needed for the calculation of the quadrupole moment of the pressure field in Equation (6). As the thermal tide is driven by large-scale mass redistribution, we then convert these surface temperature anomalies to surface pressure anomalies using the differential ideal gas law and assuming incomprehensibility ($\delta\rho = 0$) as in T. Gold & S. Soter (1969),

$$\delta p = -\Delta p \frac{\delta T}{\bar{T}} = -\Delta p \frac{T - \bar{T}}{\bar{T}}, \quad (19)$$

where $\delta p$ is the surface pressure anomaly about the mean, $p_s$, induced by the thermal tide. $\bar{T}$ is the mean temperature state, and $\delta T$ are the temperature departures from the mean (see Section 3.1). Surface temperature anomalies are then communicated to the atmosphere by atmospheric motion. Therefore, $\Delta p$ represents the fraction of the atmosphere that is thermally coupled to the surface. This is the term that encapsulates the circulation dependence of the thermal tide

$$\Delta p = \chi C_{circ} p_s, \quad (20)$$

where $C_{circ}$ is a nondimensional number describing the strength of large-scale atmospheric motion and $\chi = 0.17$ is the fraction of the atmosphere thermally coupled to the surface in the base state, taken empirically from the GCM. Therefore, to convert the temperature outputs of the simple model into pressure anomalies, we must develop a scaling for $C_{circ}$.

Following D. D. Koll & D. S. Abbot (2015), we consider the large-scale atmospheric overturning circulation as an ideal heat engine, where absorption of stellar radiation at the surface drives the motion and does work against frictional dissipation at the surface. The work done by the heat engine is given by Carnot's theorem, $W = \eta Q_{in}$, where $\eta$ is the atmosphere's thermodynamic efficiency, with $\eta = (\bar{T} - T_{eq})/\bar{T}$ and with $T_{eq}$ as the equilibrium temperature of the planet, and $Q_{in}$ is the fraction of incident stellar radiation at the surface that can be used by the atmosphere to drive circulation,

$$Q_{in} = S_o(1 - \alpha)e^{-\tau_{SW}}(1 - e^{-\tau_{LW}}). \quad (21)$$

The longwave term represents the fraction of outgoing radiation from the surface that gets absorbed by the atmosphere and is thus available to drive atmospheric motion. In the optically thin regime, most of the energy gets reradiated to space, and therefore circulation should be weak.

Energy lost as a result of friction in the boundary layer is given by (M. Bister & K. A. Emanuel 1998)

$$W = C_D \rho_s U_s^3 = C_D \frac{p_s}{R\bar{T}} U_s^3, \quad (22)$$

where $U_s$ is the average wind speed on the dayside, $R$ is the specific gas constant for the atmosphere, and $C_D$ is the surface drag coefficient.

We can therefore write an expression for the surface wind speed:

$$U_s = \left[ \frac{R \max[(\bar{T} - T_{eq}), 0]}{C_D} \frac{S_o(1-\alpha)e^{-\tau_{SW}}(1-e^{-\tau_{LW}})}{p_s} \right]^{1/3} \quad (23)$$

This expression is very similar to the scaling derived by D. D. Koll & D. S. Abbot (2015), with slight modifications to better suit the inputs of our simple model. For example, we avoid negative numbers beneath the cube root by taking the maximum of $(\bar{T} - T_{eq})$ and 0. A negative atmospheric thermodynamic efficiency occurs in our model for atmospheres that are very optically thick in the shortwave. Physically, this may correspond to an atmosphere with strong stellar absorption at high altitudes and a stable, quiescent layer at depth, which is expected to inhibit thermal tides by allowing the propagation of gravity waves (P. Auclair-Desrotour et al. 2018) and by substantially weakening the strength of the atmospheric overturning circulation (W. Kang & R. Wordsworth 2019). These cases are outside the scope of this study, as we are interested in potentially habitable rocky exoplanets and these cases are most likely to occur on planets with thick atmospheres, but have relevance to asynchronously rotating hot Jupiters (P. Auclair-Desrotour & J. Leconte 2018).

We then assume that the strength of atmospheric circulation scales with surface wind speed,

$$C_{circ} = \frac{U_s}{U_{so}}, \quad (24)$$

where $U_{so}$ is the surface wind speed in the base state described in Table 1. From Equation (23), we predict a surface wind speed of $U_{so} = 20.14 \text{ m s}^{-1}$, about twice that predicted by the GCM. The heat engine scaling idealizes the circulation of the atmosphere and should be considered as an upper limit on surface wind speeds. Overestimating wind speed will overestimate the strength of the thermal tide, so our results should be viewed as an upper limit on the ability of the thermal tide to maintain asynchronous rotation. However, because the overestimation of wind speed occurs for every case, the ratio $U_s/U_{so}$, which is what the analytic model uses, shows much better agreement with the GCM for the majority of the parameter ranges in the GCM sensitivity test. Therefore, we do not think that the overestimation of wind speed will meaningfully impact the results reported in this work.

We can combine Equations (19) and (20) to rewrite the surface pressure anomaly, $\delta p$, as

$$\delta p = -\chi \frac{U_s}{U_{so}} \frac{T - \bar{T}}{\bar{T}} p_s. \quad (25)$$

Using Equation (25), we convert temperature at each longitude from the substellar point (output from numerical integration of Equation (7)) into surface pressure anomalies. We may then perform a spherical harmonic decomposition to obtain the quadrupole term, which can be used to calculate the torque with Equation (6).

### 3.1. Approximate Analytic Solution

Equation (7) can be solved numerically; however, further insights can be gained by deriving an approximate analytic solution. Inspired by the elegance of the thermal slab model in J. Leconte et al. (2015), we linearize Equation (7) by assuming that the periodic stellar forcing and resulting surface temperature can be written as a sum of some mean state, denoted with a bar, and a small anomaly,

$$F = \overline{F} + \delta F \qquad (26)$$

$$T = \overline{T} + \delta T, \qquad (27)$$

where $F$ is equal to the full stellar forcing term. We define $\overline{F}$ as the average equatorial surface instellation,

$$\overline{F} = \frac{1}{2\pi} \int_0^{2\pi} S_o(1-\alpha) e^{-\tau_{SW}} f(\lambda) d\lambda = \frac{S_o(1-\alpha) e^{-\tau_{SW}}}{\pi}.$$

Therefore, $\delta F$ are the deviations in time from this mean stellar forcing induced by the rotation of the planet. Plugging Equations (26)–(27) into Equation (7) and expanding about small temperature anomalies,

$$C\frac{d\delta T}{dt} = \overline{F} + \delta F - \sigma_{SB}\overline{T}^4 - 4\sigma_{SB}\overline{T}^3 \delta T + F_{LW}^-(\tau_{LW}).$$

If we define $\overline{T}$ as the average surface temperature of each longitude in the perfectly distributed instellation case where $\delta F = 0$ (and including the longwave forcing of the atmosphere, which is assumed independent of rotation rate),

$$\sigma_{SB}\overline{T}^4 = \overline{F} + F_{LW}^-(\tau_{LW}),$$

then we can write the linearized equation

$$C\frac{d\delta T}{dt} = \delta F - 4\sigma_{SB}\overline{T}^3 \delta T. \qquad (28)$$

We assume that perturbations are of the form $\delta \boldsymbol{T} \propto \delta \boldsymbol{F} \propto e^{i\sigma t}$, where $\sigma$ is the forcing frequency. Noting this and rearranging Equation (28) gives the following analytic solution:

$$\delta T = \frac{\delta F}{C\omega_o} \frac{1}{1 + i\sigma/\omega_o}, \qquad (29)$$

where $\delta T$ and $\delta F$ are now the amplitudes of the perturbation for some component of the spherical harmonic decomposition of the insolation. We have defined the thermal equilibrium frequency

$$\omega_o = \frac{4\sigma_{SB}\overline{T}^3}{C}. \qquad (30)$$

Recall that we are specifically interested in the temperature anomalies associated with the quadrupole ($l = 2$) term, as this is the term that dominates the torque (A. C. Correia & J. Laskar 2003), which corresponds to $\sigma = 2(\Omega - n)$. From spherical harmonic expansion, the corresponding $\delta F = \frac{1}{6}\sqrt{\frac{15}{2\pi}} S_o(1-\alpha) e^{-\tau_{SW}}$. Plugging these expressions into Equation (19) and after some algebraic manipulation, the complex quadrupole moment is

$$\tilde{p}_2 = -\frac{1}{24}\sqrt{\frac{15}{2\pi}} \frac{S_o(1-\alpha) e^{-\tau_{SW}}}{\overline{F} + F_{LW}(\tau_{LW})} \chi p_s \frac{U_s}{U_{so}} \frac{1}{1 + i\sigma/\omega_o}$$

$$= -q_o \frac{1}{1 + i\sigma/\omega_o}, \qquad (31)$$

where $q_o$ is the amplitude of the quadrupole at zero frequency,

$$q_o = \frac{1}{24}\sqrt{\frac{15}{2\pi}} \frac{S_o(1-\alpha) e^{-\tau_{SW}}}{\overline{F} + F_{LW}(\tau_{LW})} \chi p_s \frac{U_s}{U_{so}}. \qquad (32)$$

Equation (31) has the same frequency dependence as the thermal slab model in J. Leconte et al. (2015); however, the key difference is that the parameters that encompass the effect of atmospheric composition, $q_o$ and $\omega_o$, are written analytically and therefore do not require a GCM run for each atmospheric composition, as was necessary in J. Leconte et al. (2015). The one tuning parameter in our model, $\chi$, is taken from the GCM but is considered a constant across different atmospheric types, therefore requiring only one GCM run for calibration. Therefore, this model can be used to efficiently explore how the strength of the thermal tide varies with atmospheric properties. We also note that at small forcing frequencies (corresponding to very slow rotations) the temperature anomalies about the mean become very large and the linearization assumption breaks down. Equation (31) overestimates the strength of the thermal tide by about 40% for small rotation rates compared to the numerical solution given by integrating Equation (7). Again, this overestimation means that our results should be viewed as an upper limit on the strength of the thermal tide.

The imaginary component of the complex quadrupole moment is then

$$\tilde{q} = q_o \frac{\sigma/\omega_o}{1 + \left(\frac{\sigma}{\omega_o}\right)^2}, \qquad (33)$$

the maximum of which occurs at $\sigma = \pm \omega_o$, where $\tilde{q} = \frac{1}{2}q_o$. We have defined $q_o$ as the amplitude of the thermal tide at zero frequency, where the amplitude of the quadrupole moment is given by the norm of the complex quadrupole moment,

$$|\tilde{p}_2| = q_o \frac{1}{\sqrt{1 + (\sigma/\omega_o)^2}}, \qquad (34)$$

and the sine of the lag angle is

$$\sin(2\delta) = \frac{\sigma/\omega_o}{\sqrt{1 + (\sigma/\omega_o)^2}}, \qquad (35)$$

such that $\tilde{q} = |\tilde{p}_2| \sin(2\delta)$.

Substituting Equation (33) into Equation (6) gives an analytic expression for the torque due to the thermal tide,

$$T_a = \frac{3}{2}K_a\tilde{q} = \frac{3}{2}K_a q_o \frac{\sigma/\omega_o}{1 + \left(\frac{\sigma}{\omega_o}\right)^2}. \qquad (36)$$

In Figure 3, we compare the torque ($\tilde{q}$), amplitude, and lag angle predicted by the analytic model in Equations (33)–(35) for each of the parameters of the GCM sensitivity test. The analytic model (black circles) captures the nonmonotonic behavior of the thermal tide with varying $\tau_{LW}$, with a strengthening thermal tide at low optical depths and a weakening thermal tide for large optical depths. It also skillfully predicts the sharp decline in torque with increasing shortwave optical depth. The simple model can additionally capture the near-constant value of torque as a function of ASR in this very slow rotation regime.

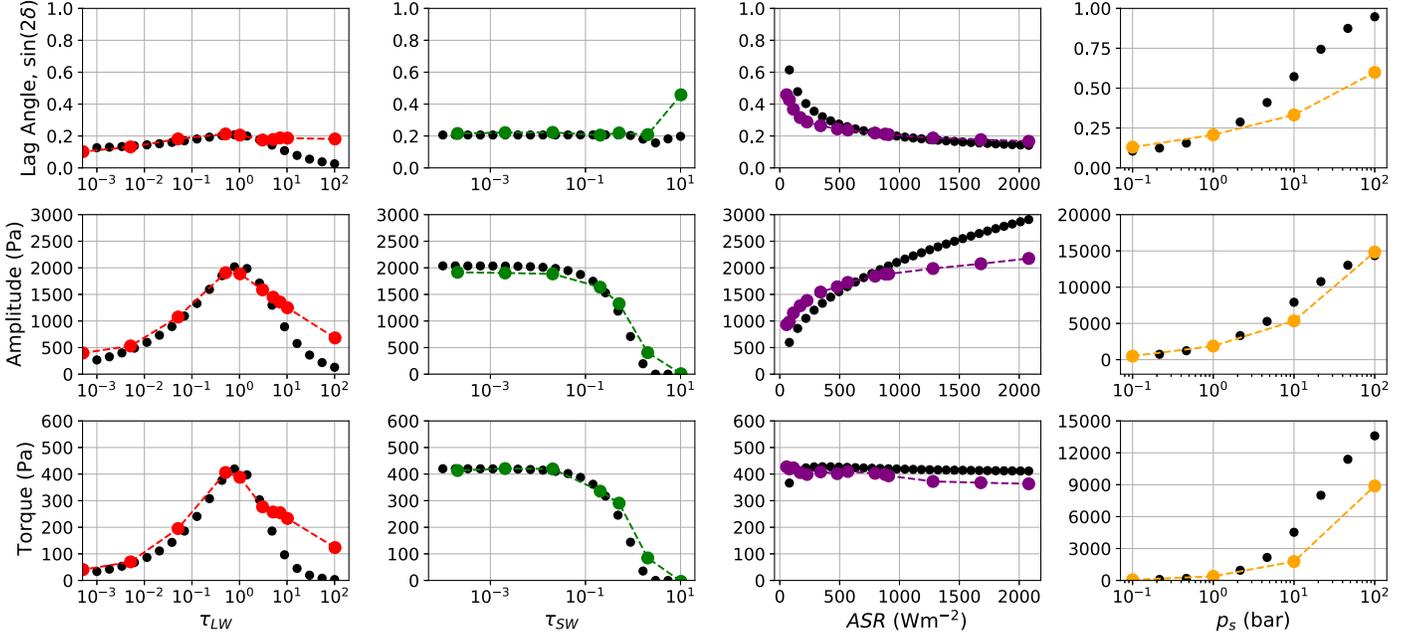

**Figure 3.** Comparison of the analytic model with the results of the targeted GCM sensitivity study. Colors indicate output from the GCM (following the color convention in Figure 2), and black circles are results from the analytic model (Equations (33)–(35)).

However, the analytic model overpredicts the strength of the thermal tide for planets with very high surface pressures by about 40% compared to the GCM, despite matching the amplitude quite well. We see that the analytic model predicts a larger lag angle compared to the GCM, which may be a limitation of the analytic model at high surface pressures. In any case, our results here demonstrate qualitatively how atmospheric composition affects the feasibility of asynchronous rotation.

It has already been shown that the thermal slab model skillfully re-creates the frequency dependence of the thermal tide (J. Leconte et al. 2015), so we do not analyze it here. In the next subsection, we develop an analytic formula for the equilibrium rotation rate of a planet undergoing gravitational and thermal tides.

### 3.2. Analytic Formula for Equilibrium Rotation Rate

The use of the constant-$Q$ model allows an analytic formulation of the equilibrium rotation rate, as in J. Leconte et al. (2015). Using the analytic formulation of the torque due to the thermal tide (Equation (36)) and the gravitational tide (Equation (4)), we write an expression for the equilibrium rotation rate where $T_g = -T_a$,

$$\frac{3}{2} K_g \, \mathrm{sgn}(2\sigma) = \frac{3}{2} K_a q_o \frac{\sigma/\omega_o}{1 + \left(\frac{\sigma}{\omega_o}\right)^2}. \tag{37}$$

In this section, we will seek only positive values of $\sigma$ (only asynchronous, prograde rotations), but we note that the solutions are completely symmetric so that if $\sigma_{eq}$ is a solution, so is $-\sigma_{eq}$. Equation (37) is rewritten as a quadratic function of $\sigma/\omega_o$,

$$\frac{K_g}{q_o K_a} \left(\frac{\sigma}{\omega_o}\right)^2 + \frac{\sigma}{\omega_o} + \frac{K_g}{q_o K_a} = 0, \tag{38}$$

which yields equilibria at forcing frequencies

$$\sigma_{eq} = \omega_o \frac{1 \pm \sqrt{1 - 4\left(\frac{K_g}{q_o K_a}\right)^2}}{2\frac{K_g}{q_o K_a}}. \tag{39}$$

Note that no real solutions are found when $\frac{K_g}{q_o K_a} > \frac{1}{2}$. This corresponds to the scenario where the thermal tide is weaker than the gravitational tide for all forcing frequencies. As noted previously, the maximum of the thermal tides occurs when $\sigma = \omega_o$. From Equation (38), this sets the criteria that asynchronous rotation can only occur when $\frac{K_g}{q_o K_a} \leqslant \frac{1}{2}$. Otherwise, we set $\sigma_{eq} = 0$.

Additionally, Equation (39) predicts two asynchronous, prograde equilibrium solutions, but only one is stable. The torque due to the thermal tide is nonmonotonic in forcing frequency, increasing with forcing frequency for $\sigma < \omega_o$ and decreasing for $\sigma > \omega_o$. The equilibrium solution where $\sigma_{eq} < \omega_o$ is therefore unstable, because a small perturbation toward a faster (slower) rotation rate will lead to a stronger (weaker) thermal tide torque, which will further increase (decrease) the rotation rate. On the other hand, the equilibrium solution where $\sigma_{eq} > \omega_o$ is stable because a small increase (decrease) in rotation rate will lead to a weakened (strengthened) torque, which will cause the planet's rotation rate to decrease (increase), returning the rotation rate back to the equilibrium. We therefore consider only equilibria where $\sigma_{eq} \geqslant \omega_o$, the larger root. In conclusion, the equilibrium forcing frequency

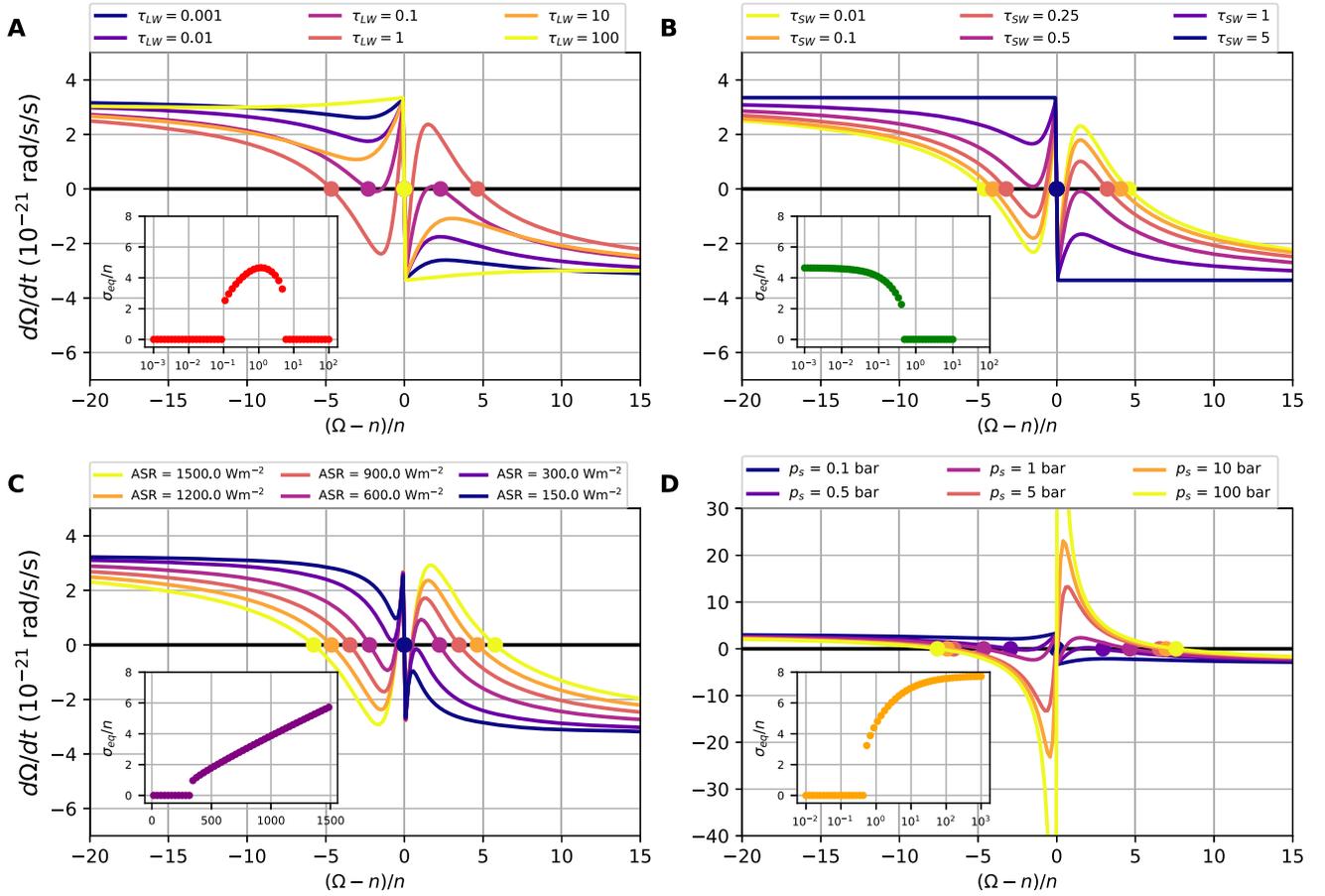

**Figure 4.** Demonstration of how atmospheric parameters determine a planet's equilibrium rotation rate, (a) $\tau_{\mathrm{LW}}$, (b) $\tau_{\mathrm{SW}}$, (c) ASR, and (d) $p_s$. All panels assume that the planet orbits a 0.5 $M_\odot$ star at 0.33 au. Unless stated, atmospheric parameters are set at their base value. The time derivative of the spin rate is calculated using Equation (1) and the analytic model of the torque due to the thermal tide. Filled circles mark the equilibrium forcing frequency, calculated using Equation (40). The inset plots in each panel show the equilibrium rotation rate ($\sigma_{\mathrm{eq}}/n$) as a function of each variable.

(and therefore rotation rate) is given by

$$\sigma_{\mathrm{eq}} = 2(\Omega_{\mathrm{eq}} - n) = \begin{cases} \omega_o \dfrac{1 + \sqrt{1 - 4\left(\frac{K_g}{q_o K_a}\right)^2}}{2\frac{K_g}{q_o K_a}} & \dfrac{K_g}{q_o K_a} \leqslant \dfrac{1}{2} \\ 0 & \dfrac{K_g}{q_o K_a} > \dfrac{1}{2}. \end{cases}$$
(40)

Though this formulation has been used before (e.g., A. C. M. Correia & J. Laskar 2001; J. Leconte et al. 2015), in the past the key parameters controlling the strength of the thermal tide, $q_o$ and $\omega_o$, relied on input from a GCM. Our analytic formulae for $q_o$ (Equation (32)) and $\omega_o$ (Equation (30)) as a function of atmospheric properties allow a novel exploration of how the equilibrium rotation rate of a planet depends on atmospheric properties.

In the following section, we utilize the analytic model of the thermal tides to explore the parameter regime under which asynchronous rotation is possible.

## 4. Results

In Figure 4, we plot the time derivative of the spin rate, $d\Omega/dt = (T_g + T_a)/I$, as a function of forcing frequency for each of the targeted sensitivity tests (using the analytic model), assuming that the planet orbits a 0.5 $M_\odot$ star at 0.33 au. We demonstrate here that a planet's equilibrium rotation rate depends strongly on its atmospheric composition. For example, in Figure 4(a) the analytic model predicts that the planet will be tidally locked for atmospheres with $\tau_{\mathrm{LW}} < 0.1$ and $\tau_{\mathrm{LW}} > 8$. Similarly in Figure 4(b), planets with $\tau_{\mathrm{SW}} \geqslant 0.5$ will have synchronous rotations.

For all atmospheric properties, the equilibrium rotation rate does not smoothly decrease toward a tidally locked state. Instead, each jumps abruptly from some asynchronous state into a tidally locked state. This effect comes from the nonmonotonic dependence of the torque on the forcing frequency. From Equation (40), as $K_g/q_o K_a$ approaches $1/2$, the equilibrium rotation rate approaches $\omega_o$, not zero. This is due to the fact that the maximum in the thermal tide occurs at $\sigma = \omega_o$ and there are no stable equilibria for $\sigma < \omega_o$.

An especially interesting case is shown in Figure 4(c), where the analytic model shows that a planet's equilibrium rotation rate decreases as ASR decreases. This initially seems in contradiction with the results of the GCM study in Figure 2, which showed a near-constant torque as a function of ASR owing to a complete compensation between the increase in amplitude of the thermal tide and the decrease in lag angle with increased ASR. However, the GCM sensitivity tests were run at a constant, small forcing frequency $\sigma = n$, corresponding to a planet in a 2:1 orbital resonance. In Figure 4(c), we see that in the limit of small forcing frequencies changes to ASR result in similar net torques because of the compensating effect of the amplitude and lag angle. We can understand this clearly by examining the ASR dependence of Equation (33) in the limit $\sigma \ll \omega_o$ (small forcing frequency). In this limit, the torque is approximately linear in $\sigma$ with $\tilde{q} \approx (q_o/\omega_o)\sigma$. From Equation (33), because $\overline{F}$ and $F_{\mathrm{LW}}^-(\tau_{\mathrm{LW}})$ both contain an ASR term, the only ASR dependence of $q_o$ is in the strength of circulation, which has an $\mathrm{ASR}^{1/3}$ dependence. Assuming that the thermal inertia of the atmosphere is much larger than the surface (about a factor of 3 for a 1-bar atmosphere), $\omega_o \propto \mathrm{ASR}^{5/12}$. Therefore, the ASR dependence of the torque in the slow-rotation regime is $\mathrm{ASR}^{-1/12}$, a very weak dependence. This is shown in both the GCM and simple model results in Figure 3 (third row, third column), where the torque has a very small decrease with increasing ASR. In the regime where $\sigma \gg \omega_o$, however, $\tilde{q} \propto q_o\omega_o(1/\sigma) \propto \mathrm{ASR}^{3/4}$. Therefore, the equilibrium rotation rate of a planet is expected to decrease with decreasing ASR, as seen in Figure 4(c).

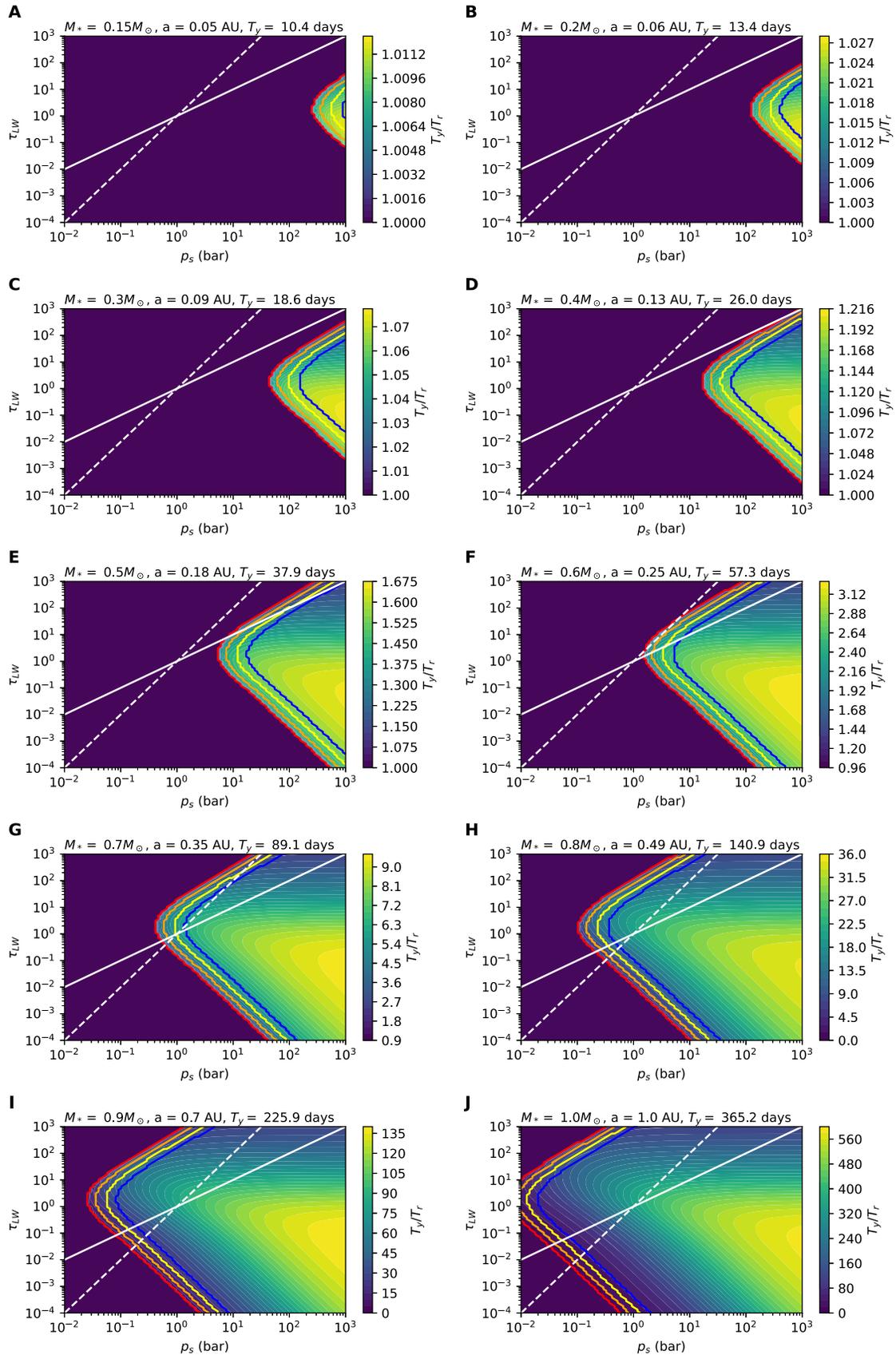

**Figure 5.** Equilibrium rotation rate (as a ratio of the period of a year, $T_y$, and the period of one rotation, $T_r$, both in Earth days) of an Earth-mass planet calculated using Equation (40) as a function of $\tau_{LW}$ and $p_s$. Purple regions correspond to synchronous rotation. For each stellar mass, we calculate the equilibrium rotation at an orbital semimajor axis, $a$, such that the incoming stellar radiation is equal to 1360 W m$^{-2}$. The dashed white line shows the path through the parameter space when $\tau_{LW} \propto p_s^2$, and the solid white line shows $\tau_{LW} \propto p_s$. The colored contour lines show how the boundary between synchronous (purple) and asynchronous rotation varies with the albedo of the planet: 0.3 (red), 0.5 (orange), 0.7 (yellow), 0.9 (blue). In all cases, $\tau_{SW} = 0.1$.

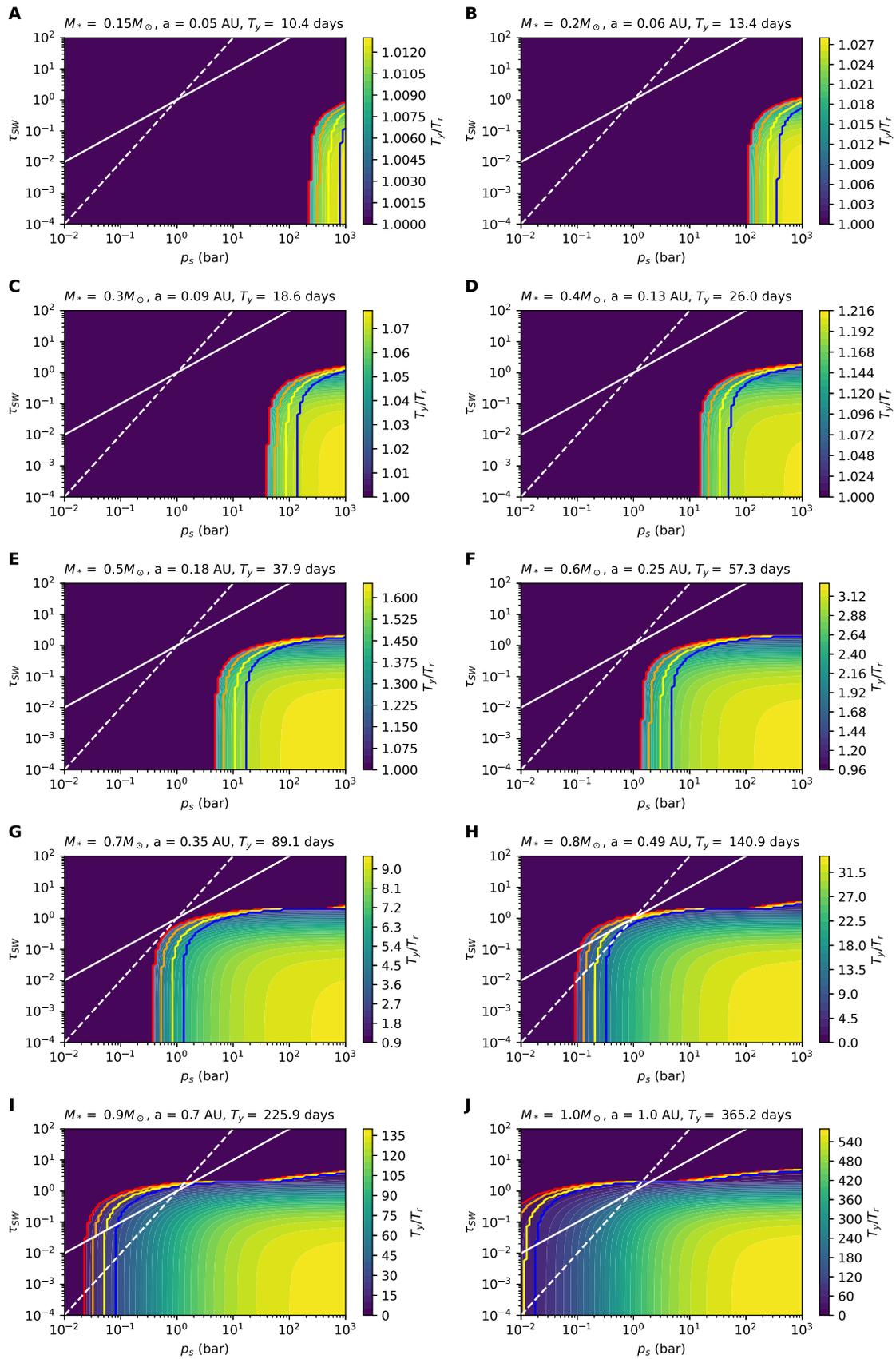

**Figure 6.** Same as in Figure 5, but for $\tau_{\mathrm{SW}}$ and $p_s$. In all cases, $\tau_{\mathrm{LW}} = 1$.

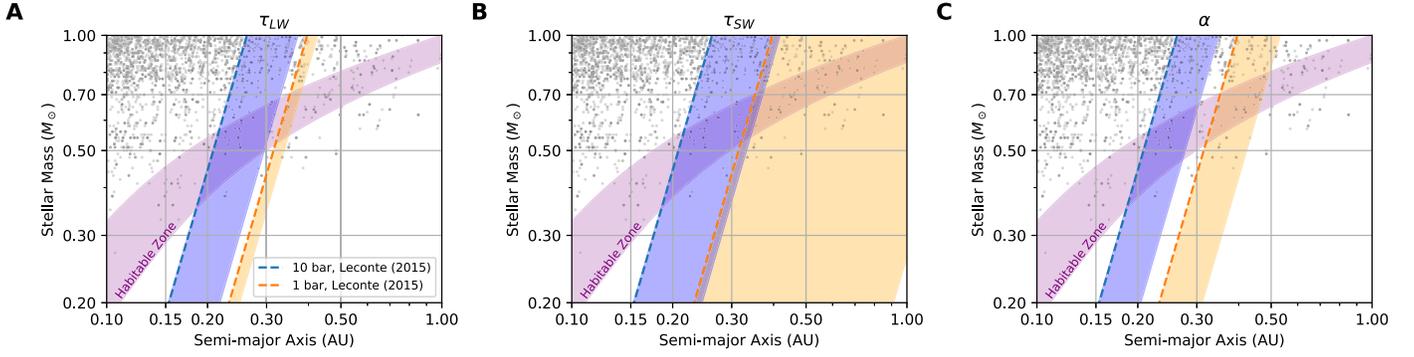

**Figure 7.** Critical semimajor axis, $a_c$, for asynchronous rotation for varying atmospheric parameters. In all panels, the results from J. Leconte et al. (2015) are shown for the 1-bar (orange) and 10-bar (blue) cases. The shading represents how the critical semimajor axis shifts as you vary (a) longwave optical depth (between 5 and 50 for 10 bars and between 1 and 10 for 1 bar), (b) shortwave optical depth (between 0.6 and 2 for 10 bars and 1 bar), and (c) planetary albedo (between 0.2 and 0.9 for 10 bars and 1 bar). Gray circles present confirmed exoplanets between 0.5 and 3 Earth radii from the NASA Exoplanet Archive.

This result suggests a positive climate feedback, wherein as a planet's rotation rate is slowed, it may enter a regime where convective clouds are formed over the substellar point (M. J. Way et al. 2018). This would increase planetary albedo, decreasing ASR and weakening the thermal tide, which would further decreasing the planetary rotation rate. Future work could examine the conditions under which such a feedback may operate and whether it can plummet a planet into a synchronous rotation state.

With the combined results of the GCM and the analytic model, we have explored how each of these atmospheric parameters can affect the equilibrium rotation rate of a planet. In the next section, we use the analytic model to map out the regions in the parameter space where asynchronous rotation is possible.

In the GCM sensitivity tests, we varied one parameter at a time, which allowed us to construct an analytic theory for how the strength of the thermal tide depends on various atmospheric properties. We now utilize the analytic model to explore the response of the thermal tide to changes in atmospheric properties that may be expected to vary together. We begin with the coevolution of surface pressure and longwave/shortwave optical depth.

J. Leconte et al. (2015) predicted a maximum in the thermal tide with increasing surface pressure due to the corresponding increase in stellar absorption at higher layers in the atmosphere. We extend this idea in this work by observing that atmospheres that are optically thick in the longwave have weak thermal tides owing to the mitigation of temperature gradients between the dayside and nightside.

In Figure 5, we plot the equilibrium rotation rate of a planet as a function of $p_s$ and $\tau_{\mathrm{LW}}$. As the equilibrium rotation rate is a function of orbital parameters and atmospheric properties (Equation (40)), we explore this parameter space for many star types, restricting the calculation to planets receiving an Earth-like stellar flux, $S = 1360 \, \mathrm{W \, m^{-2}}$, to represent the approximate location of the HZ (J. F. Kasting et al. 1993; R. K. Kopparapu et al. 2013). This directly defines the orbital semimajor axis, $a$, by assuming that the stellar luminosity varies with stellar mass (J. Scalo et al. 2007),

$$L_* = L_\odot 10^{4.101\mu^3 + 8.162\mu^2 + 7.108\mu}, \tag{41}$$

where $\mu = \log(M_*/M_\odot)$. The orbital period, and therefore the orbital mean motion $n$, is calculated from Kepler's third law.

In Figure 5, the purple regions indicate the values of $p_s$ and $\tau_{\mathrm{LW}}$ for which the planet is likely to be tidally locked. For small stars (e.g., Figure 5(a)), the HZ is extremely close to the star, leading to a very strong gravitational tide. Even for atmospheres up to 1000 bars, the simple model predicts that asynchronous rotation (denoted by nonpurple colors) is extremely unlikely. For larger stars, the HZ extends farther from the star as a result of the increase in stellar luminosity, which increases the number of atmospheric states, which could lead to asynchronous rotation. In general, high-pressure atmospheres that are relatively optically thin provide the strongest thermal tides, but in general optical thickness should scale with surface pressure. We plot examples of this relationship using white lines, which show the optical depth if it is assumed to increase linearly with surface pressure (solid, no pressure broadening) and as the square of surface pressure (dashed, with pressure broadening). This provides intuition for where a typical atmosphere on a rocky exoplanet may fall in this parameter space. For example, an atmosphere far to the right of the white lines represents a very high pressure atmosphere that is also very transparent, which may not be expected for most atmospheric compositions. Therefore, though there are atmospheric states around $0.5 \, M_\odot$ or smaller stars that this model predicts to have asynchronous rotation, these states may not be very likely.

In the colored contours, we demonstrate how the boundary between synchronous and asynchronous rotation shifts with planetary albedo. As albedo increases (red to blue), the boundary shifts to higher surface pressures, further limiting the parameter regimes under which asynchronous rotation is possible. The change in albedo can be considered as a change in cloud cover or an enhancement of Rayleigh scattering with surface pressure.

In Figure 6, we perform the same analysis as a function of $\tau_{SW}$ and $p_s$. The region of asynchronous rotation is strongly capped by $\tau_{SW}$, with essentially no solutions of asynchronous rotation occurring around any star for atmospheres with $\tau_{SW} > 2$. Increasing the shortwave optical depth weakens the thermal tide both by limiting the amount of the stellar radiation that can be absorbed at the surface and by inhibiting large-scale atmospheric motion. Only planets with thick atmospheres that are extremely transparent to stellar radiation are predicted to escape tidal locking in this model. Though not explicitly explored here, the redshifted spectra of M stars may therefore present further limitations on the feasibility of asynchronous rotation for their HZ planets, depending on the relative impacts of enhanced stellar absorption and decreased Rayleigh scattering.

For stars smaller than about 0.5 $M_{\odot}$ (M stars), asynchronous rotation in the HZ may therefore be very unlikely, as it requires atmospheres with high surface pressures but relatively low shortwave and longwave optical depths and low planetary albedo.

For the most direct comparison with the GCM results of J. Leconte et al. (2015), and to contextualize our results with the known population of terrestrial exoplanets, we calculate the critical semimajor axis beyond which a planet would have a synchronous rotation following J. Leconte et al. (2015). We define the critical semimajor axis, $a_c$, as the smallest semimajor axis a planet with some thermal tide could have while maintaining asynchronous rotation, where $\frac{K_g}{q_o K_a} = \frac{1}{2}$. Assuming that $q_o$ depends only weakly on $a$ within the HZ, the critical semimajor axis is

$$a_c = \left(\frac{10\pi}{3}\right)^{1/6} \left(\frac{GM_* R_p^2 k_2 \overline{p}}{q_o Q}\right)^{1/3}, \qquad (42)$$

identical to Equation (2) in J. Leconte et al. (2015). From Equation (42), a stronger thermal tide (larger $q_o$) decreases $a_c$, such that the planet could be closer to the star and still maintain asynchronous rotation.

In Figure 7, we expand on the results of J. Leconte et al. (2015) by demonstrating how atmospheric composition shifts the critical semimajor axis and limits the likelihood of asynchronous rotation for HZ planets. The purple shaded region indicates the bounds of the HZ as a function of stellar mass using Equation (41) and assuming that the inner and outer edges of the HZ occur at an incident stellar flux of 1.05 and 0.34 $F_\oplus$, respectively (R. K. Kopparapu et al. 2013). Gray circles represent confirmed exoplanets with radii between 0.5 and 3 Earth radii as reported by the NASA Exoplanet Archive.

We plot the results of J. Leconte et al. (2015, their Figure 3) using Equation (42) by plugging in their values of $q_o$ for a 1-bar and 10-bar $N_2-CO_2$ atmosphere with $\alpha = 0.2$ and stellar flux of 1366 W m$^{-2}$ (reported in their Table 1). To obtain the same $q_o$ at each pressure level, our analytic model requires $\tau_{LW}$ and $\tau_{SW}$ to be equal to 5 and 0.5 in the 10-bar case and 1 and 0.58 in the 1-bar case (though different combinations of the two could give the same result). The shading in each panel demonstrates how $a_c$ shifts if the value of each parameter is increased independently. As expected from our previous results, an increase in $\tau_{LW}$, $\tau_{SW}$, or $\alpha$ weakens the thermal tide (and therefore $q_o$), therefore requiring the planet to be farther from the star to maintain asynchronous rotation. For planets orbiting stars less than about 0.5 $M_{\odot}$, planets within the HZ may or may not be tidally locked depending on the characteristics of the atmosphere.

## 5. Discussion

In this work, we used a hierarchical modeling approach to study the dependence of the strength of the thermal tide on the atmospheric composition of rocky exoplanets. Our model suggests the following conclusions:

1. The torque due to the thermal tide is nonmonotonic with increasing longwave optical depth ($\tau_{LW}$) owing to the competing effects of reduced zonal temperature gradients and strengthening atmospheric circulation.
2. The thermal tide weakens rapidly with increased shortwave optical depth ($\tau_{SW}$) owing to the joint effect of limiting shortwave radiation at the surface and inhibiting atmospheric circulation by stabilizing the atmosphere.
3. The equilibrium rotation rate of a planet decreases with increased planetary albedo, suggesting interesting climate−rotation feedbacks. Additionally, the presence of haze could weaken the thermal tide and preclude asynchronous rotation.
4. For HZ planets orbiting stars smaller than 0.5 $M_{\odot}$, thick atmospheres may be unable to maintain asynchronous rotation owing to the corresponding increase in atmospheric opacity.

The analytic model employed here was useful for its computational efficiency and interpretability, but the formulation of the model required many simplifying assumptions. To simplify the radiative transfer in the atmosphere, we employed the WTG approximation such that the atmospheric temperature profile was assumed horizontally uniform. This assumption would tend to break down for HZ planets around very small stars, as even near synchronicity the rotation rate of the planet may be rapid enough to allow horizontal temperature gradients to appear. By employing the WTG approximation in these cases, we will underestimate the amplitude of surface temperature gradients, which will underestimate the torque. For example, underestimating the torque by a factor of two would decrease the critical semimajor axis by about 20% (Equation (42)), which may shift our results to predict asynchronous rotation around lower-mass stars for some atmospheric states. We have also ignored all boundary layer dynamics, especially temperature inversions on the nightside and turbulent heat fluxes, which may similarly lead to an overestimation of nightside temperature and therefore an underestimation of the torque. Conversely, the representation of atmospheric circulation as a heat engine tends to overestimate surface wind speeds, which may lead to an overestimate of the torque due to the thermal tide. Though the ratio $U_s/U_{so}$, which is what is actually used by the analytic model to calculate the torque, shows better agreement with the GCM than $U_s$ alone, in some cases it can deviate by about 50%, which would lead to about a 10% overestimation of the critical semimajor axis. Therefore, each of these assumptions will have some, and often an opposing, effect on the quantitative strength of the thermal tide, but we believe that these uncertainties are small compared to, say, the uncertainty in the quality factor, $Q$. An order-of-magnitude variation in $Q$ would shift the critical semimajor axis by about 50%. More rigorous quantification of the magnitude of the thermal tide should be explored using more complex models.

This work is especially relevant given the push to characterize the atmospheres of rocky planets using JWST (J. Ih et al. 2023; J. Lustig-Yaeger et al. 2023; S. Zieba et al. 2023; R. Hu et al. 2024). Measurements of the thermal emission spectra upon secondary eclipse are used to detect the presence of an atmosphere on an assumed tidally locked planet by measuring the emission temperature of the dayside. However, if a planet is observed to have an asynchronous rotation, for example, through phase-resolved emission spectroscopy (K. B. Stevenson et al. 2014), despite its location within the tidal locking radius of the star, perhaps the observed rotation rate may be used as an additional constraint on the atmospheric composition of the planet using the theory of the atmospheric thermal tide.

Finally, the dependence of the rotation rate of a planet on the characteristics of its atmosphere may present interesting and previously unexplored climate−rotation feedbacks. One example may be a dayside cloud feedback, wherein a decrease in the rotation rate of the planet leads to enhanced dayside cloud cover (J. Yang & D. S. Abbot 2014; M. J. Way et al. 2018), increasing the albedo of the planet and therefore weakening the thermal tide, which further decreases the rotation rate. Another example may be a tipping point into a synchronous rotation, as a decreasing rotation rate leads to nightside temperatures cold enough to cause atmospheric condensation and collapse. However, as synchronous rotation is an unstable equilibrium of the thermal tide when using the constant-$Q$ model, the formation of a secondary atmosphere may allow a planet to escape tidal locking even if it becomes tidally locked at some point in its history. The atmospheric and orbital evolution of a planet may therefore be intimately coupled, and this relationship should be considered in studies of the evolution of habitable worlds.

## Acknowledgments

This research has made use of the NASA Exoplanet Archive, which is operated by the California Institute of Technology, under contract with the National Aeronautics and Space Administration under the Exoplanet Exploration Program. R.W. acknowledges funding from NSF award AST-1847120 (CAREER) and the Leverhulme Center for Life in the Universe. We also thank two anonymous reviewers, whose helpful comments improved this work.

All processed GCM data and code needed to reproduce the figures in this paper are publicly available on GitHub: https://github.com/andreamsalazar/ExoTides_SW2024. The processed GCM data are also available in Zenodo at doi:10.5281/zenodo.13381086 (Salazar 2024). Raw GCM output can be made available upon request.

## ORCID iDs

Andrea M. Salazar  https://orcid.org/0000-0002-3861-9592
Robin Wordsworth  https://orcid.org/0000-0003-1127-8334